\documentclass[reprint,superscriptaddress,showpacs,amsmath,amssymb,prb,twocolumn]{revtex4}

\usepackage{mathtools}
\usepackage{natbib}
\usepackage{siunitx}
\usepackage{array}
\usepackage{xcolor}
\usepackage{subfigure} 
\usepackage{amsmath}
\usepackage{amsfonts}
\usepackage{amssymb}
\usepackage{graphicx}
\usepackage{hyperref}

\providecommand{\vect}[1]{{\boldsymbol{#1}}}

\DeclareMathOperator{\sech}{sech}

\begin{document}

\title{Dzyaloshinskii-Moriya Induced Spin-Transfer Torques in Kagome Antiferromagnets}

\author{Davi R.~Rodrigues}
\affiliation{Department of Electrical and Information Engineering, Polytechnic University of Bari, 70125 Bari, Italy}
\author{Akshaykumar~Salimath}
\affiliation{Department of Engineering Sciences, University of Agder, 4879 Grimstad, Norway}
\author{Karin~Everschor-Sitte}
\affiliation{Faculty of Physics and Center for Nanointegration Duisburg- Essen (CENIDE), University of Duisburg-Essen, 47057 Duisburg, Germany}
\author{Kjetil M. D.~Hals}
\affiliation{Department of Engineering Sciences, University of Agder, 4879 Grimstad, Norway}
\date{\today}
 
\begin{abstract}
In recent years antiferromagnets (AFMs) have become very promising for nanoscale spintronic applications due to their unique properties such as THz dynamics and absence of stray fields. Manipulating antiferromagnetic textures is currently, however, limited to very few exceptional material symmetry classes allowing for staggered torques on the magnetic sublattices. In this work, we predict for kagome AFMs with broken mirror symmetry a new coupling mechanism between antiferromagnetic domain walls (DWs) and spin currents, produced by the relativistic Dzyaloshinskii-Moriya interaction (DMI). We microscopically derive the DMI's free energy contribution for the kagome AFMs. Unlike ferromagnets and collinear AFMs, the DMI does not lead to terms linear in the spatial derivatives but instead renormalizes the spin-wave stiffness and anisotropy energies. Importantly, we show that the DMI induces a highly nontrivial twisted DW profile that is controllable via two linearly independent components of the spin accumulation. This texture manipulation mechanism goes beyond the concept of staggered torques and implies a higher degree of tunability for the current-driven DW motion compared to conventional ferromagnets and collinear AFMs. 

\end{abstract}
\pacs{}

\maketitle

\begin{figure}[t] 
	\centering 
	\includegraphics[scale=0.30]{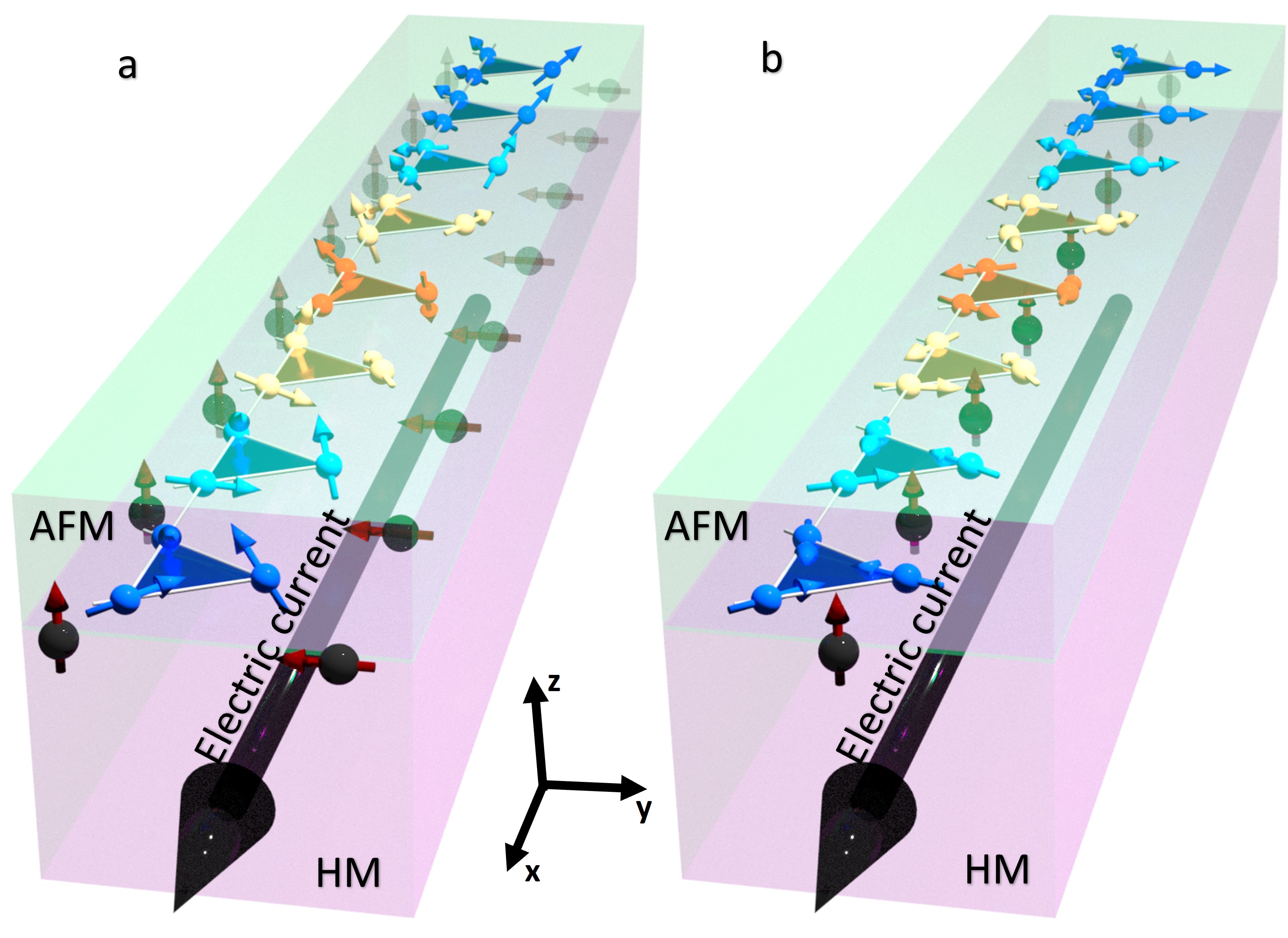}  
	\caption{(color online). 
		{\bf a}.~Current-driven DW motion in kagome AFMs with DMI and broken mirror symmetry.
		Due to the DMI, the DW attains a twisted out-of-plane profile, which couples the spin texture to both the in-plane and out-of-plane spin components of the injected spin current.  
		{\bf b}.~Current-driven DW motion in kagome AFMs without DMI. 
		The spins are confined to rotate in the plane defined by the AFM layer, and the DW only couples to the out-of-plane spin component of the injected spin current. The color code of the magnetization represents $\cos^2\theta$, where $\theta$ represents the rotation along the $z$-axis. In this representation, parallel/antiparallel ground states have the same color (blue) and the DW is emphasized.
	}
	\label{Fig1} 
\end{figure}

The manipulation of magnetic textures by electrical currents was a breakthrough discovery in spintronics and galvanized a new generation of non-volatile memory devices with high performance and low power requirements~\cite{Parkin2008,Bader2010,Zutic2004}. Further advantages come from the use of antiferromagnetic materials, allowing for more compact devices and increased performance speeds~\cite{Jungwirth2016,Baltz2018}. 
The manipulation and observation of antiferromagnetic textures, however, still present a great obstacle. The electric control of the magnetization is contingent on the crystal layering structure and symmetries~\cite{Chen2014,Gomonay2016,Shiino2016,Zelezny2018}. So far, current-driven manipulation has only been observed in antiferromagnets (AFMs) 
having a crystal symmetry allowing for a staggered torque on the magnetic sublattices~\cite{Zelezny2014, Chen2014, Wadley2016, Bodnar2018,Kimata2019}. 

A particularly interesting class of antiferromagnetic materials is the noncollinear antiferromagnets (NCAFMs)~\cite{Andreev1980}. 
Unlike ferromagnets and collinear AFMs, whose spin orders are characterized by a single vector field, the NCAFMs have a SO(3)-valued order parameter field. Consequently, the NCAFMs are expected to exhibit richer and more complex spintronic properties than those observed in ferromagnets and collinear AFMs. 

Recently, it has been demonstrated that laser pulses can scan and write domain walls (DWs) in NCAFMs~\cite{Reichlova2019}. Astonishingly, their speed and direction of motion can be controlled by a single spin-wave source with tunable frequency~\cite{Rodrigues2021}. However, only a few works have addressed the current-induced dynamics of the NCAFMs~\cite{Gomonay2012, Tserkovnyak2017a, Ochoa2018, Yamane2019, Li2021,Lund2021}, and little is known about how relativistic interactions influence the shape and current-driven control of the DWs. Specifically, it is expected that several NCAFMs could have a significant Dzyaloshinskii-Moriya interaction (DMI). For example, in NCAFMs with kagome structure, it has been shown that the broken mirror symmetry of the kagome plane yields a complex DMI that varies on the atomic scale~\cite{Wills2001,Elhajal2002,Elhajal2005,Yildirim2006,Zorko2008}. 
As in ferromagnets and collinear AFMs~\cite{Emori2013, Ryu2013, Schulz2015,Pan2018,Qaiumzadeh2018}, it is anticipated that the NCAFMs' DMI is crucial in understanding the equilibrium and spintronic properties of the DWs~\cite{Liu2017,Li2019,Sugimoto2020}.

Here, we demonstrate that in kagome AFMs with broken mirror symmetry, the DMI induces a distinctly twisted DW shape, which differs markedly from any observed spin textures in ferromagnets and collinear AFMs. We investigate how these nontrivial DWs are manipulable by spin currents injected into the thin-film kagome AFM from an adjacent heavy metal layer (see Fig.~\ref{Fig1}). We find that these twisted DWs can be controlled by a novel spin-transfer torque (STT) mechanism that couples the DW’s center-of-mass coordinate additionally to spin currents polarized parallel to the kagome plane. In contrast, for mirror-symmetric kagome planes, the DWs are only affected by the out-of-plane component of the spin accumulation. Consequently, the DWs in kagome AFMs with broken mirror symmetry couple to two linearly independent components of the spin current, allowing for a high degree of tunability in the electrical manipulation of the DWs. Furthermore, the new STT mechanism opens the door for detecting the kagome AFMs' complex DMI via the current-driven dynamics of the DWs. 

\begin{figure}[t] 
\centering 
\includegraphics[scale=1.0]{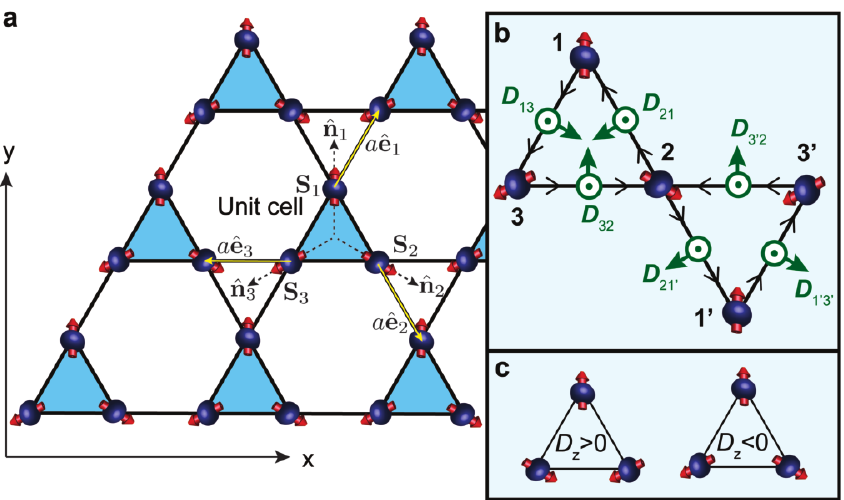}  
\caption{(color online). 
{\bf a}. Illustration of the lattice vectors $\hat{\vect{e}}_{i}$ and in-plane easy axes $\hat{\vect{n}}_{i}$. 
{\bf b}. The green arrows show the DMI vectors on the bondings for the  case where $D_z >0$ and $D_{\parallel} > 0$. 
The spins in $H_D$ should be summed counter-clockwise around the triangles (indicated by the black arrows on the bondings). 
{\bf c}. Two 120$^\circ$ spin configurations of opposite chirality. 
For clarity, all figures are illustrated in the limit $D_{\parallel} \rightarrow 0$, in which the out-of-plane tilting becomes negligible.  
}
\label{Fig2} 
\end{figure}

We model the kagome AFM with DMI in the exchange approximation, where the isotropic exchange energy is assumed to be much stronger than the relativistic interactions produced by the spin-orbit coupling. 
In this case, the mutual orientation of the sublattice spins is only weakly affected by the creation of spin textures and their dynamics. 
The spin Hamiltonian of the system is 
\begin{equation}\label{Eq:Hamiltonian}
H = H_{e} + H_{D} + H_{a},
\end{equation}
where $H_{e}= J\sum_{\langle ij \rangle} \vect{S}_i\cdot\vect{S}_j$ is the isotropic exchange interaction between the neighboring lattice sites $\langle ij \rangle$,  $H_{D} = \sum_{\langle ij \rangle} \vect{D}_{ij}\cdot\left(\vect{S}_{i}\times\vect{S}_{j}\right)$ represents the DMI, and
 $H_{a}=  \sum_{i} [ K_z \left( \vect{S}_{i}\cdot\hat{\vect{z}} \right)^2 - K\left( \vect{S}_{i}\cdot\hat{\vect{n}}_{i} \right)^2]$
describes the easy plane ($K_z>0$) and easy axes ($K>0$) anisotropy energies. 
The unit vectors $\hat{\vect{n}}_i$ refer to the in-plane easy axis at lattice site $i$. The kagome NCAFM can be divided into three spin sublattices with in-plane easy axes $\hat{\vect{n}}_1 = [0,1,0]$, $\hat{\vect{n}}_2 = [\sqrt{3}/2,-1/2,0]$, and $\hat{\vect{n}}_3 = [-\sqrt{3}/2,-1/2,0]$, respectively.  The unit vectors connecting the three sublattices are $\hat{\vect{e}}_1 = [1/2,\sqrt{3}/2,0]$, $\hat{\vect{e}}_2 = [1/2,-\sqrt{3}/2,0]$, and $\hat{\vect{e}}_3 = [-1,0,0]$, and $a$ is the lattice constant (see Fig.~\ref{Fig2}a). 

The structure of the DMI is determined by the symmetry of the system. 
The bulk material consists of stacked kagome lattice layers. In such three-dimensional kagome lattice, the inversion symmetry is broken, as the layers are typically shifted among each other. However, each kagome lattice layer still corresponds to a mirror plane. In this case, the DMI vectors $D_{ij}$ are confined to be along the $z$-axis~\cite{Elhajal2002,Elhajal2005}.
Often, however, the mirror symmetry of the kagome lattice is broken, e.g., due to the presence of nonmagnetic atoms between the kagome planes such as in jarosites~\cite{Wills2001,Elhajal2002,Elhajal2005}. The symmetry can also be lowered in heterostructures by sandwiching the kagome AFM between two different materials. In these cases, the DMI vectors additionally have an in-plane component~\cite{Wills2001,Elhajal2002,Elhajal2005,Yildirim2006,Zorko2008,Liu2017} (Fig.~\ref{Fig2}b). Thus, the most general form of the DMI within a unit cell is
\begin{subequations}
\begin{align}
\vect{D}_{21} =& D_z \hat{\vect{z}} + D_{\parallel} ( \hat{\vect{e}}_2\times \hat{\vect{z}} ), \label{Eq:DMIvectorA} \\
\vect{D}_{13} =& D_z \hat{\vect{z}} + D_{\parallel} ( \hat{\vect{e}}_1\times \hat{\vect{z}} ), \label{Eq:DMIvectorB} \\
\vect{D}_{32} =& D_z \hat{\vect{z}} + D_{\parallel} ( \hat{\vect{e}}_3\times \hat{\vect{z}} ). \label{Eq:DMIvectorC}
\end{align}
\end{subequations}
The DMI vectors on the bondings connecting the unit cells are determined by inversion about the site of sublattice $2$, i.e.\ $D_{21'}=D_{21}, D_{1'3'}=D_{13}$ and $D_{3'2}=D_{32}$ (Fig.~\ref{Fig2}b).
The DMI vectors in Eqs.~\eqref{Eq:DMIvectorA}-\eqref{Eq:DMIvectorC} have two important implications for the ground state spin configurations of Eq.~\eqref{Eq:Hamiltonian}. 
First, the sign of $D_z$ competes with the easy-axis anisotropy to select a 120$^\circ$ ordering of the sublattice spins with a fixed chirality (Fig.~\ref{Fig2}c). For $D_z=0$, these two configurations correspond to an energy minimum of the exchange interaction even though they have different easy-axis anisotropy contributions~\cite{Yamane2019}.
Second, the in-plane DMI leads to a weak out-of-plane tilting of the spins: $S_{i,z}\propto -{\rm sign} [D_{\parallel}]$. 
Consequently, $D_{\parallel}$ yields a weak ferromagnetic state.   
Below, we consider $D_z < 0$ such that the Hamiltonian~\eqref{Eq:Hamiltonian} has two ground-state configurations, in which the in-plane components $\vect{S}_{i, \parallel}$ of the spins are oriented along 
$\pm \hat{\vect{n}}_i$. 

To derive a free energy functional, which provides a coarse-grained description of the Hamiltonian~\eqref{Eq:Hamiltonian},
we write the three sublattice spins of a unit cell as~\cite{Dombre1989}
\begin{equation}\label{Eq:Representation}
\vect{S}_{\iota} (t) = \frac{S \mathcal{\mathbf{R}} (t) \left[ \hat{\vect{n}}_{\iota} + a\vect{L} (t) \right] }{ \| \hat{\vect{n}}_{\iota} + a\vect{L} (t)  \| } , \ \  \iota \in \{ 1,2,3 \} .
\end{equation}
Here, $\iota = 1,2,3$ labels the sublattices within a unit cell, $\mathcal{\mathbf{R}}$ is a rigid rotation matrix that represents the order parameter of the NCAFM, whereas $\vect{L}$ is a vector that describes an overall tilting of the spins away from their equilibrium direction. 
Because $\mathcal{\mathbf{R}}$ and $\vect{L}$ are defined by six independent parameters, the above representation yields a complete description of all possible configurations of the three sublattice spins.
Note that $\mathcal{\mathbf{R}}$ and  $\vect{L}$ are constant within a unit cell and vary smoothly on length scales comparable to the system's exchange length $\lambda = a\sqrt{J/K}$. Furthermore, the exchange approximation implies that the tilting $a\vect{L}$ is small. Substituting Eq.~\eqref{Eq:Representation} into Eq.~\eqref{Eq:Hamiltonian}, and expanding the Hamiltonian to second order in $a\vect{L}$ and the spatial gradients of $\mathcal{\mathbf{R}}$, yields in the continuum limit $a\rightarrow 0$ the free energy functional    
\begin{equation}
	F[\mathcal{\mathbf{R}}, \vect{m}] =  \int dA \left[ \mathcal{F}_R(\mathcal{\mathbf{R}}) + \mathcal{F}_m(\vect{m}) \right], \label{Eq:FreeEMR}
\end{equation}
where $dA=dx dy$.
The free energy densities 
\begin{subequations}\label{Eq:F1}
\begin{align}
	\mathcal{F}_R =&  \mathcal{J}^{\alpha\beta}_{ijkl} \partial_{\alpha}R_{ij} \partial_{\beta} R_{kl}   +  \mathcal{K}_{ijkl} R_{ij}  R_{kl}  , \label{Eq:FR2} \\
	\mathcal{F}_m =& a_2\vect{m}^2  + \eta_0  m_{z}^2 - \vect{h}_{D}\cdot\vect{m},  \label{Eq:Fm2} 
\end{align}
\end{subequations}
originate respectively from the rotation matrix $\mathcal{\mathbf{R}}$ and the vector field $\vect{m}= \mathbf{T}\vect{L}$ with $T_{\alpha\beta} = \delta_{\alpha\beta} + (1/3)\sum_{\iota} n_{\iota,\alpha}n_{\iota,\beta}$  and $\delta_{\alpha\beta}$ being the Kronecker delta. Throughout, Einstein’s summation convention is implied for repeated indices. In Eq.~\eqref{Eq:FR2},
\begin{equation}
\mathcal{J}^{\alpha\beta}_{ijkl}= \Lambda_{jl}^{\alpha\beta}\delta_{ik} + \mathcal{D}^{\alpha\beta}_{ijkl},
\end{equation}
parametrizes the spin-wave stiffness of the kagome AFM. Its first contribution,
\begin{equation}
\Lambda_{jl}^{\alpha\beta} = -\Lambda_0[ \mathcal{A}_{jl}^{\alpha\beta} + \mathcal{B}_{jl}^{\alpha\beta} +  \mathcal{C}_{jl}^{\alpha\beta} ],
\end{equation}
represents the isotropic exchange interaction. Here, $\Lambda_{0}= (4S^{2}J)/\sqrt{3}$ is a constant and the tensors $\mathcal{A}$, $\mathcal{B}$, and 
$\mathcal{C}$ are determined by the easy axes and lattice vectors via the relationships
$\mathcal{A}_{jl}^{\alpha\beta}= n_{1,j}n_{3,l}e_{1,\alpha}e_{1,\beta}$, $\mathcal{B}_{jl}^{\alpha\beta}= n_{2,j}n_{1,l}e_{2,\alpha}e_{2,\beta}$, and
$\mathcal{C}_{jl}^{\alpha\beta}= n_{3,j}n_{2,l}e_{3,\alpha}e_{3,\beta}$.
The second contribution to the spin stiffness,
\begin{equation}
\mathcal{D}^{\alpha\beta}_{ijkl} = -\epsilon_{\tau ik} [ \tilde{D}_{13,\tau} \mathcal{A}_{jl}^{\alpha\beta} + \tilde{D}_{21,\tau} \mathcal{B}_{jl}^{\alpha\beta} + \tilde{D}_{32,\tau} \mathcal{C}_{jl}^{\alpha\beta} ],
\end{equation}
is induced by the DMI. Here, $\epsilon_{\tau ik}$ is the Levi-Civita symbol and $\tilde{\vect{D}}_{ij}= (4S^2/\sqrt{3})\vect{D}_{ij}$ are the DMI vectors in the continuum limit. 
In Eq.~\eqref{Eq:FR2}, the tensor
\begin{equation}\label{eq:TensorK}
\mathcal{K}_{ijkl}= \kappa_{ijkl} + d_{ijkl}
\end{equation}
describing the anisotropy of the NCAFM also has two contributions. The first one,
\begin{equation}
\kappa_{ijkl} = \sum_{\iota}[\tilde{K}_{z}n_{\iota, j}n_{\iota, l}\delta_{zi}\delta_{zk}-\tilde{K}n_{\iota, i}n_{\iota, j}n_{\iota, k}n_{\iota, l}],
\end{equation}
is linked to the anisotropy energies, where $\tilde{K}_{z}= 4S^2K_{z}/a^{2}\sqrt{3} $ and $\tilde{K}= 4S^2K/a^{2}\sqrt{3}$ are the anisotropy constants in the continuum limit. The second contribution to the anisotropy of the NCAFM, Eq.~\eqref{eq:TensorK},
\begin{equation}
d_{ijkl} = \frac{2\epsilon_{\tau ik}}{a^2}  [ \tilde{D}_{13,\tau}\mathcal{M}_{jl} + \tilde{D}_{21,\tau} \mathcal{N}_{jl}  + \tilde{D}_{32,\tau} \mathcal{O}_{jl}  ],
\end{equation}
is determined by the DMI with the tensors 
$\mathcal{M}_{jl}= n_{1,j}n_{3,l} $, $\mathcal{N}_{jl}= n_{2,j}n_{1,l}$, $\mathcal{O}_{jl}= n_{3,j}n_{2,l}$.
In Eq.~\eqref{Eq:Fm2}, the constants are $a_{2}=36S^{2}J/\sqrt{3}$ and  $\eta_0 = 12K_{z}S^{2}/\sqrt{3}$.
Furthermore, $\vect{h}_D= -(24S^2D_{\parallel}/a) \hat{\vect{z}}$ is the field induced by the in-plane DMI, which leads to weak ferromagnetism. 
A derivation of Eq.~\eqref{Eq:FreeEMR} is provided in App.~\ref{Sec:Appendix1}.

The free energy~\eqref{Eq:FreeEMR} captures the long-wavelength physics of kagome AFMs with DMI, and is the first key result of this Letter. 
To our knowledge, there has been no microscopic derivation of the DMI's contribution to the free energy functional of kagome AFMs.
The free energy of kagome AFMs without DMI was microscopically derived in Refs.~\onlinecite{Ulloa2016, Rodrigues2021}.
Interestingly, Eq.~\eqref{Eq:FR2} reveals that the DMI does not produce any terms that are linear in the spatial derivatives of the order parameter field $\mathcal{\mathbf{R}}$. Instead, it renormalizes the spin-wave stiffness and anisotropy energy via the highly anisotropic tensors $\mathcal{D}^{\alpha\beta}_{ijkl}$ and $d_{ijkl}$.
This differs significantly from DMI contributions in ferromagnets and collinear AFMs. 
For example, unlike ferromagnets and collinear AFMs, kagome AFMs described by Eq.~\eqref{Eq:FreeEMR} do not support helical spin structures as ground states.

In the following, we study the physics of a DW connecting two domains being in the two energetically degenerate ground states of a kagome AFMs with broken mirror symmetry (see Fig.~\ref{Fig1}a). 
To this end, we examine a spin texture along the $x$-axis and parameterize the rotation matrix by nautical angles: 
\begin{equation}
\mathcal{\mathbf{R}}(x) = \mathcal{\mathbf{R}}_z (\theta (x)) \mathcal{\mathbf{R}}_y (\phi (x)) \mathcal{\mathbf{R}}_x (\psi (x)).\label{Eq:R}
\end{equation}
Here, $\theta$, $\phi$ and $\psi$ represent rotations about the $z$-, $y$- and $x$-axis, respectively. 
Furthermore, we consider a parameter regime in which the easy-plane anisotropy is the dominant interaction produced by the spin-orbit coupling: $K/K_z \ll 1$, $\| \vect{D}_{ij}\| /K_z \ll 1$.
This is consistent with the typical material parameters of kagome AFMs~\cite{Sakuma2003, Matan2006, Szunyogh2009, Park2018a}.

In the absence of DMI, a DW corresponds to a rotation $\mathcal{\mathbf{R}}(x)=\mathcal{\mathbf{R}}_z (\theta (x))$, where $\psi=\phi=0$, and $\theta (x)$ varies smoothly between $0$ and $\pm \pi$ in the DW region.
Because the DMI produces a highly anisotropic form of the spin-wave stiffness $\mathcal{J}^{\alpha\beta}_{ijkl}$, it may lead to finite out-of-plane rotations  $\phi (x)$ and $\psi (x)$  of the DW.
We expect these rotations to be small and linear in the DMI (to leading order). 
Below, we therefore keep terms up to second order in $\phi $ and $\psi $ in the free energy density $\mathcal{F}_R (\theta, \phi, \psi )$ and solve the equilibrium equations to first order in the DMI. 
The expression for $\mathcal{F}_R (\theta, \phi, \psi )$ is given in App.~\ref{Sec:Appendix1}.
The equilibrium equations for the nautical angles and $\vect{m}$ are found from the
variational equations $\delta F/\delta\vartheta = 0$ ($\vartheta\in \{ \theta,\phi, \psi \}$) and $\delta F/\delta\vect{m}= \boldsymbol{0}$.

From Eq.~\eqref{Eq:FreeEMR}, we notice that the equilibrium value of $\vect{m}$ is not affected by the underlying spin texture and is a function only of the in-plane DMI contribution:
\begin{equation}
	\vect{m}= -\frac{D_{\parallel}}{a \sqrt{3} J}\left(\frac{1}{1+ \eta_0/a_2}\right)\hat{\vect{z}}.
\end{equation}
This allows to integrate out the field $\vect{m}$ in the effective action description of the DW.
Note that $\eta_0/a_2\sim K_z/J \ll 1$ and, thus, $\vect{m}\approx  -(D_{\parallel}/a\sqrt{3}J)\, \hat{\vect{z}}$ for a static kagome AFM.  

For the angle $\theta$, we obtain the equilibrium equation
\begin{equation}
\partial_x^2 \theta = \frac{2\tilde{K}}{\Lambda_0 - \sqrt{3} \tilde{D}_z } \sin (2\theta) , \label{Eq:thetaEq}
\end{equation} 
which has the well known solution 
\begin{equation}
\theta (x) = \sigma_1 2\arctan [\exp (\sigma_2 (x-r)/\lambda_{dw})] .\label{Eq:theta}
\end{equation} 
Here, $\lambda_{dw}= \sqrt{(\Lambda_0-\sqrt{3} \tilde{D}_z )/4\tilde{K}}$ is the DW width, $r$ is the center of the DW, whereas
$\sigma_1 \in \{1,-1\}$ and $\sigma_2 \in \{1,-1\}$ are determined by the boundary conditions of $\theta$. 
Notice that the DMI-induced change of the DW width in Eq.~\eqref{Eq:theta} resembles the ferromagnetic case~\cite{Tretiakov2010}.

Varying the free energy~\eqref{Eq:FreeEMR} with respect to the nautical angle $\psi$ produces the following equilibrium equation
\begin{equation}
\psi = \frac{\tilde{K}\phi \sin(2\theta)}{2(\hat{K}_z + \tilde{K}\cos^2\theta )} - \frac{\sqrt{3}\tilde{D}_{\parallel} \partial_x [(\partial_x\theta) \sin\theta]}{8(\hat{K}_z + \tilde{K}\cos^2\theta )} , \label{Eq:psiEq}
\end{equation}
where $\hat{K}_z= \tilde{K}_z - \sqrt{3} \tilde{D}_z  / a^2$. Thus, $\psi$ is completely determined by the solutions of the angles $\theta$ and $\phi$.
The functional form of the nautical angle $\phi$ is dictated by 
\begin{equation}
\epsilon^2 \partial_{\tilde{x}}^2 \phi (\tilde{x}) - Q(\tilde{x}) \phi (\tilde{x}) + R(\tilde{x}) = 0, \label{Eq:phiEq}
\end{equation}
which has the form of an inhomogeneous Schr\"odinger equation.
Here, we have introduced the dimensionless coordinate $\tilde{x}= (x-r)/\lambda_{dw}$ and the constant  $\epsilon^2= \Lambda_0/2\hat{K}_z\lambda_{dw}^2$. The functions $Q$ and $R$ are defined by
\begin{subequations}
\begin{align}
Q (\tilde{x}) =& 1 + k \left[ 1 - 3 \sech^2 (\tilde{x}) \right] , \\
R (\tilde{x}) =& -\sigma_1 \alpha_{\parallel} \left[  \sech (\tilde{x}) - 2\sech^3 (\tilde{x}) \right] ,
\end{align}
\end{subequations}
with constants $k= \tilde{K}/\hat{K}_z$ and $\alpha_{\parallel}= \sqrt{3}\tilde{D}_{\parallel}/8\hat{K}_z \lambda_{dw}^2$.
An exact analytical solution of Eq.~\eqref{Eq:phiEq} is not known.
However, from the expression of $\lambda_{dw}$, we see that $\epsilon \sim \sqrt{k} \ll 1$.
Therefore, Eq.~\eqref{Eq:phiEq} is on a form suitable for the Wentzel–Kramers–Brillouin (WKB) approximation~\cite{Bender1999}, which leads to 
$\phi(\tilde{x}) = \int_{-\infty}^{\infty}d\tilde{x}'  G(\tilde{x},\tilde{x}') R(\tilde{x}')$ with $G(\tilde{x},\tilde{x}')= \exp [-(1/\epsilon) |\int_{\tilde{x}'}^{\tilde{x}} dt \sqrt{Q(t)}|]/2\epsilon (Q(\tilde{x})Q(\tilde{x}'))^{1/4}$.
In the limits $k\ll 1$, an asymptotic expansion of the above integral implies the approximate solution
\begin{equation}
\phi(\tilde{x})= - \sigma_1 \alpha_{\parallel} \left[  \sech (\tilde{x}) - 2\sech^3 (\tilde{x}) \right] . \label{Eq:phi}
\end{equation} 
Substituting Eq.~\eqref{Eq:phi} and Eq.~\eqref{Eq:theta} into Eq.~\eqref{Eq:psiEq}, yields the following expression for the nautical angle $\psi$
\begin{equation}
\psi(\tilde{x})=  \sigma_2 2\alpha_{\parallel}  \sech^2 (\tilde{x})\tanh (\tilde{x})  . \label{Eq:psi}
\end{equation} 
Eqs.~\eqref{Eq:theta}, \eqref{Eq:phi} and \eqref{Eq:psi} are the second key result of this Letter, and determine the DW profile of kagome AFMs.
Importantly, we notice that the DMI forces the DW to develop a highly nontrivial out-of-plane twist state when the mirror symmetry of the kagome lattice is broken. 

Next, we will investigate how this nontrivial DW texture couples to a spatial uniform spin accumulation $\vect{\mu}_s$. 
The source of this spin accumulation does not play a role in the results we derive. However, in an experiment, 
$\vect{\mu}_s$ typically originates from the spin Hall effect (SHE) in an adjacent heavy metal layer, which generates a spin current into the AFM, as sketched in Fig.~\ref{Fig1}.
An effective action $\mathcal{S}_{\mathrm{eff}}= \int dt dA\,   \mathcal{L}_{\mathrm{eff}}$ of the kagome AFM is characterised by the Lagrangian density 
\begin{equation}
\mathcal{L}_{\mathrm{eff}} = \frac{m_{\mathrm{eff}}}{4} {\rm Tr} [\dot{\mathbf{R}}^{T}  \dot{\mathbf{R}}] - \mathcal{F}_s  - \mathcal{F}_R . \label{Eq:L}
\end{equation}
Here,  the first term represents the kinetic energy where $m_{\mathrm{eff}} = 2\hbar^2 / \sqrt{3}J a^2$. 
Additionally, we have included $\mathcal{F}_s$, which captures the free energy contributions from the spin accumulation and the DMI:
$\mathcal{F}_s=  (\hbar/12aSJ) ( h_{D, i} - g_r \mu_{s,i} )\epsilon_{ijk} [\mathbf{R}^T \dot{\mathbf{R}}]_{jk}$. Here, $g_r$ parametrizes the reactive STT.
The free energy density $\mathcal{F}_R$ is given by Eq.~\eqref{Eq:FR2}.   

The dissipative processes of the spin system are governed by the dissipation functional
\begin{equation}
\mathcal{G} = \int dt dA \left[  \frac{\alpha_d}{8} {\rm Tr} [\dot{\mathbf{R}}^{T} \dot{\mathbf{R}} ] + \frac{g_d}{2} \mu_{s,i} \epsilon_{ijk} [\mathbf{R}\dot{\mathbf{R}}^T]_{jk} \right]   , \label{Eq:G}
\end{equation}
where $\alpha_{d}$ and $g_d$ are parameters controlling the damping and dissipative STT, respectively. 
Eqs.~\eqref{Eq:L}-\eqref{Eq:G} are microscopically derived in App.~\ref{Sec:Appendix2}.

In the stationary regime, the STT-driven motion of the DW is well described by the center-of-mass coordinate $r$. 
The time evolution of $r$ can be found by substituting the WKB solutions Eqs.~\eqref{Eq:theta}, \eqref{Eq:phi} and \eqref{Eq:psi} into Eqs.~\eqref{Eq:L} and \eqref{Eq:G} and integrating over the spatial coordinates.
We then find the following action and dissipation functionals: 
\begin{align}
\mathcal{S} =& L_y \int dt  \frac{m_{\mathrm{eff}} \dot{r}^2}{\lambda_{dw}} , \nonumber \\
\mathcal{G} =& L_y \int dt \left[ \frac{\alpha_d \dot{r}^2}{2\lambda_{dw}}  - \sigma_1 \sigma_2\pi g_d \left( \frac{\alpha_{\parallel}}{2}\mu_{s,y} +  \mu_{s,z} \right) \dot{r} \right]  . \nonumber
\end{align}
Here, $L_y$ is the width of the sample.
Note that $\mathcal{F}_s$ in Eq.~\eqref{Eq:L} does not contribute to the dynamics of $r$, because $\vect{h}_D$ and $\vect{\mu}_s$ are static. 
Consequently, the field-like STT does not influence the DW motion, and the main driving torque is the damping-like STT, similar to the collinear AFM case~\cite{Shiino2016}. 
From the variational equation $\delta\mathcal{S}/\delta r = \delta\mathcal{G}/\delta \dot{r}$, we find the equation of motion for $r$
\begin{equation}
\frac{2m_{\mathrm{eff}}}{\lambda_{dw}} \ddot{r} = - \frac{\alpha_d}{\lambda_{dw}} \dot{r} + \sigma_1\sigma_2\pi g_d\left( \frac{\alpha_{\parallel}}{2}\mu_{s,y} +  \mu_{s,z} \right) . 
\end{equation}
In the stationary regime, in which $\ddot{r} \rightarrow 0$, the DW approaches the terminal velocity
\begin{equation}
v_{dw}= \frac{\sigma_1 \sigma_2\pi g_d\lambda_{dw}}{\alpha_d}\left(\frac{a^2\sqrt{3}\tilde{D}_{\parallel}}{16 \left(a^2 \tilde{K}_z - \sqrt{3} \tilde{D}_z\right) \lambda_{dw}^2 } \hat{\vect{y}} + \hat{\vect{z}}\right)\cdot{\vect{\mu}}_{s} . \label{Eq:Vdw}
\end{equation}
The current-driven DW velocity~\eqref{Eq:Vdw} is the third key result of this Letter and demonstrates that the DMI produces a novel force acting on the DWs via $x$ polarized spin currents.
In absence of DMI, the DWs only couple to the $z$-component of $\vect{\mu}_s$~\cite{Yamane2019}, i.e., through the first term of Eq.~\eqref{Eq:Vdw}.
Consequently, the DMI enables manipulation of the DWs via two linearly independent components of $\vect{\mu}_s$.
Furthermore, Eq.~\eqref{Eq:Vdw} opens the possibility to probe the in-plane DMI via measurements of the current-driven DW motion.
For example, for a DW driven by $\vect{\mu}_s || \hat{\boldsymbol{y}}$, the terminal velocity $v_{dw}$ provides a measure of $\tilde{D}_{\parallel}$.

To summarize, we have microscopically derived the DMI’s free energy contribution for kagome AFMs. We have shown that the DMI  renormalises the spin-wave stiffness and anisotropy energies, but does not lead to the typical terms linear in the spatial derivatives. In kagome AFMs, a non-vanishing in-plane DMI can be observed in jarosites and heterostructures by juxtaposing non-magnetic materials with the kagome planes. Furthermore, we have investigated how the DMI influences the shape and current-driven motion of  DWs in kagome AFMs. 
Our findings reveal that the DMI causes the DWs to develop a twisted spatial profile. A major consequence of the emerging twist state is that it gives rise to a new STT, which enables the control of DWs via two linearly independent torques. Importantly, this suggests that NCAFMs are particularly attractive for use in spintronic devices as they offer a higher degree of control of the electrically operated DWs.

This work received funding from the Research Council of Norway via Grant No. 286889 and was supported by the DFG (German Research Foundation) via the Emmy Noether project 320163632 and the TRR 173 -- 268565370 (project B12). D.R. acknowledges funding by the Ministry of Education, University and Research of Italy via the PON Project “NGS–New Satellites Generation Components” (COD.ID. ARS01\_01215).

\appendix

\section{Derivation of free energy}\label{Sec:Appendix1}
In what follows, we consider a spin system described by the microscopic Hamiltonian~\eqref{Eq:Hamiltonian} and derive an expression for the free energy functional in the continuum limit (assuming the temperature $T\rightarrow 0$).
To this end, we first derive the energy contribution from one unit cell (labeled by the index $u$) for $H_e$, $H_D$, and $H_a$ in Eq.~\eqref{Eq:Hamiltonian} and express the spins in terms of $\vect{L}$ and $\mathbf{R}$ 
using Eq.~\eqref{Eq:Representation}. Secondly, we sum over all the unit cells and take the 
continuum limit by converting the sum to an integral via $\sum_u \rightarrow \int dA/a_c$.  Here, $a_c= a^2\sqrt{3}/4$ is the area of the the kagome lattice's unit cell.
Throughout, we keep terms to second order in $a\boldsymbol{L}$ and the spatial gradients of $\mathbf{R}$.

For $|a\vect{L}| \ll 1$, the spin representation~\eqref{Eq:Representation} becomes
\begin{equation}\label{S_approx}
	\boldsymbol{S}_{\iota}\approx S\mathbf{R}[\hat{\boldsymbol{n}}_{\iota} + a(\boldsymbol{L}-(\hat{\boldsymbol{n}}_{\iota}\cdot\boldsymbol{L})\hat{\boldsymbol{n}}_{\iota} ] .
\end{equation}
The net spin polarization of a unit cell is  $\boldsymbol{S}_{tot}=\sum_{\iota=1}^{3}\boldsymbol{S}_{\iota}=3aS\mathbf{R}\vect{m}$, where $T_{\alpha\beta}=\delta_{\alpha\beta} - (1/3)\sum_{\iota=1}^{3} n_{\iota,\alpha}n_{\iota,\beta}$ and $\vect{m}= \mathbf{T}\boldsymbol{L}$. For the considered kagome AFM, the operator $\mathbf{T}$ is diagonal with elements $2T_{xx}=2T_{yy}=T_{zz}=1$.

The energy contribution of unit cell $u$ originating from the Heisenberg exchange term $H_e=J\sum_{\langle ij\rangle}\boldsymbol{S}_{i}\cdot\boldsymbol{S}_{j}$ is
\begin{align}\label{exchange}
	H_{e,u} =& J[\boldsymbol{S}^{l}_{1}\cdot(\boldsymbol{S}^{l+\hat{e}_{1}}_{3}+\boldsymbol{S}^{l-\hat{e}_{1}}_{3})\notag\\
	&+\boldsymbol{S}^{l}_{2}\cdot(\boldsymbol{S}^{l+\hat{e}_{2}}_{1}+ 
	\boldsymbol{S}^{\l-\hat{e}_{2}}_{1})+\boldsymbol{S}^{l}_{3}\cdot(\boldsymbol{S}^{l+\hat{e}_{3}}_{2}+\boldsymbol{S}^{l-\hat{e}_{3}}_{2})].
\end{align}
Here, $l$ denotes the position of the spin within unit cell $u$ and $l\pm \hat{e}_{1}$ is the nearest neighbor lattice site connected to $\l$ via the lattice vector $\pm a\hat{\boldsymbol{e}}_i$. 
The spatial variations of the the spin $\boldsymbol{S}^{l\pm\hat{e}_{i}}_{j}$ is captured by 
the gradient expansion $\boldsymbol{S}^{l\pm\hat{e}_{i}}_{j}\approx\boldsymbol{S}^{l}_{j}\pm a(\hat{\boldsymbol{e}}_{i}\cdot\boldsymbol{\nabla})\boldsymbol{S}_{j}^{l}+\frac{a^{2}}{2}(\hat{\boldsymbol{e}}_{i}\cdot\boldsymbol{\nabla})^{2}\boldsymbol{S}_{j}^{l}$.  
Substituting this expansion along with the expression \eqref{S_approx} into Eq.~\eqref{exchange} produces the energy contribution
\begin{equation}
	H_{e,u}=9a^{2}S^{2}J\vect{m}^2+ a_c \Lambda_{ij}^{\alpha\beta}[\partial_{\alpha}\mathbf{R}^{T}\partial_{\beta}\mathbf{R}]_{ij},
\end{equation}
where the tensor $\Lambda_{ij}^{\alpha\beta}$ is defined in the main text.

Similarly, the DMI energy of unit cell $u$ can be expressed as 
\begin{align}\label{dmi}
	H_{D,u} =& \vect{D}_{13} \cdot [\boldsymbol{S}^{l}_{1}\times (\boldsymbol{S}^{l+\hat{e}_{1}}_{3}+\boldsymbol{S}^{l-\hat{e}_{1}}_{3}) ] \notag\\
	&+	\vect{D}_{21} \cdot [  \boldsymbol{S}^{l}_{2}\times(\boldsymbol{S}^{l+\hat{e}_{2}}_{1}+  \boldsymbol{S}^{\l-\hat{e}_{2}}_{1}) ]\notag\\
	&+ \vect{D}_{32} \cdot[ \boldsymbol{S}^{l}_{3}\times(\boldsymbol{S}^{l+\hat{e}_{3}}_{2}+\boldsymbol{S}^{l-\hat{e}_{3}}_{2}) ].
\end{align}
Upon substitution of Eq.~\eqref{S_approx} and the gradient expansion of the spins, Eq.~\eqref{dmi} yields the DMI energy
\begin{equation}
	H_{D,u} = a_c [ \mathcal{D}^{\alpha\beta}_{ijkl} \partial_{\alpha}R_{ij} \partial_{\beta} R_{kl} + d_{ijkl} R_{ij}  R_{kl} - h_{D,i}m_i ], \label{Eq:Fd} 
\end{equation}
where  $\mathcal{D}^{\alpha\beta}_{ijkl}$, $d_{ijkl}$, and $\vect{h}_D$ are given in the main text. 

We split the anisotropy energy into easy axes and easy plane contributions: $H_{a,u}= H^{(\textrm{axes})}_{a,u} + H^{(\textrm {plane})}_{a,u}$.
The easy axes anisotropy energy of unit cell $u$ is $H^{(\textrm{axes})}_{a,u}=-\sum_{\iota=1}^3 K(\boldsymbol{S}_{\iota}\cdot\hat{\boldsymbol{n}}_{\iota})^{2}$, which to second order in the out-of-equilibrium quantities becomes (again, using Eq.~\eqref{S_approx})  
\begin{equation}
	H^{(\textrm {axes})}_{a,u}= -KS^{2}\sum_{\iota=1}^3(n_{\iota, i}n_{\iota, j}n_{\iota, k}n_{\iota, l})(R_{ij}R_{kl}) . 
\end{equation}
Correspondingly, we find for the easy plane anisotropy $H^{(\textrm {plane})}_{a,u}=\sum_{\iota=1}^3 K_{\iota}(\hat{\boldsymbol{z}}\cdot\boldsymbol{S}_{\iota})^2$ the energy contribution
\begin{equation}
	H^{(\textrm {plane})}_{a,u}=K_{z}S^{2}\sum_{\iota=1}^3 n_{\iota, i}n_{\iota, j}R_{zi}R_{zj}+3a^{2}S^{2}K_{z}(\hat{\boldsymbol{z}}\cdot \vect{m} )^{2}.
\end{equation}

Thus, the energy of unit cell $u$ is $H_u= H_{e,u} + H_{D,u} + H_{a,u}$  and the total energy of the spin system becomes $H= \sum_u H_{u}$.
The free energy functional~\eqref{Eq:FreeEMR} is obtained by taking the continuum limit $\sum_{u} \rightarrow\int dxdy/a_c$, and grouping the terms involving $\mathbf{R}$ and $\mathbf{m}$ into $\mathcal{F}_R$ and $\mathcal{F}_m$, respectively.  

Parametrizing the rotation matrix in terms of nautical angles using Eq.~\eqref{Eq:R},  $\mathcal{F}_R$ in Eq.~\eqref{Eq:FR2}  can to second order in $\phi$ and $\psi$ be written as
\begin{align}
	\mathcal{F}_R =& \frac{3\Lambda_0}{4} \left[ (1-\phi^2) (\partial_x \theta)^2  + (\partial_x \phi)^2  \right]  - 3 \tilde{K}\cos^2\theta\notag\\
	&  + \frac{3}{2} \left[ (\hat{K}_z + \tilde{K}\cos^2\theta) (\phi^2 + \psi^2)  - \tilde{K}\phi\psi \sin(2\theta)  \right]\notag \\
	& +  \frac{3\sqrt{3}}{4}  \tilde{D}_z  (\partial_x\theta)^2\notag\\
	 &- \frac{3\sqrt{3}}{8} \tilde{D}_{\parallel} \left[ (\partial_x\phi)\cos\theta + (\partial_x\psi)\sin\theta \right] (\partial_x\theta) . \label{Eq:FR} 
\end{align}

\section{Action and dissipation functionals }\label{Sec:Appendix2}
\subsection{The action of the kagome AFM}
The action of the spin system is $\mathcal{S}= \int {\rm dt} [\mathcal{T} - H - H_s] $. 
Here, the first term is the kinetic energy $\mathcal{T}= \sum_i \hbar \vect{A}(\vect{S}_i)\cdot \dot{\vect{S}}_i $, where $\vect{A}$ is vector potential satisfying $\vect\nabla \times \vect{A}(\vect{S}_i)   = \vect{S}_i/S$, $H$ is the Hamiltonian~\eqref{Eq:Hamiltonian} of the isolated spin system, and
$H_s=\sum_{i}  \lambda_{r}\boldsymbol{\mu}_{s}\cdot\boldsymbol{S}_{i}$ describes the interaction energy between the spins and the spin accumulation $\boldsymbol{\mu}_s$.

The kinetic energy of unit cell $u$ is $\mathcal{T}_u=\sum_{\iota=1}^3\hbar A_{\alpha}[\boldsymbol{S}_{\iota}] \dot{{S}}_{\iota,\alpha}$.
Expanding the vector potential $\boldsymbol{A}(\boldsymbol{S}_{k})$ to first order in the out-of-equilibrium quantities and using Eq.~\eqref{S_approx}, yields the expression
\begin{align}\label{Eq:Texp}
	\sum_{\iota=1}^3 \hbar A_{\alpha}[\boldsymbol{S}_{\iota}]   \dot{S}_{{\iota},\alpha} \approx& \sum_{\iota=1}^3\hbar S[A_{\alpha}(\mathbf{R}\hat{\boldsymbol{n}}_{\iota})\cdot (\dot{ \mathbf{R}}\hat{\boldsymbol{n}}_{\iota})_{\alpha}\notag\\
	&+a\epsilon_{\alpha\beta\gamma}L_{\alpha}n_{\iota,\beta}(\mathbf{R}^{T} \dot{ \mathbf{R}}\hat{\boldsymbol{n}}_{\iota})_{\gamma}] ,
\end{align}
where we have used the relationship $\epsilon_{\alpha\beta\gamma}\partial_{\beta}A_{\gamma}= S_{\alpha}/S$ as well as the property $\epsilon_{\alpha\beta\gamma}\mathbf{R}_{\alpha\alpha'}\mathbf{R}_{\beta\beta'}\mathbf{R}_{\gamma\gamma'}=\epsilon_{\alpha'\beta'\gamma'}$ of the rotation matrix. 
The first term in Eq.~\eqref{Eq:Texp} can be disregarded as it is a topological term that does not influence the equations of motion of the rotation matrix.  
The quantity $\mathbf{R}^{T}\dot{ \mathbf{R}}$ is antisymmetric and can thus be written as  $(\mathbf{R}^{T}\dot{ \mathbf{R}})_{ij}=-\epsilon_{ij\alpha}\mathcal{V}_{\alpha}$, where $\mathcal{V}_{x}$, $\mathcal{V}_{y}$, and $\mathcal{V}_{z}$ represent the three independent tensor elements.       
Substituting this expression into Eq.~\eqref{Eq:Texp}, we find the following kinetic energy of one unit cell $\mathcal{T}_u=3a\hbar S\vect{m}\cdot\boldsymbol{\mathcal{V}}$.  

Next, we consider the coupling $H_s$ to the spin accumulation $\boldsymbol{\mu}_s$, which is assumed to be spatial uniform. 
Using that $\sum_{\iota=1}^{3}\boldsymbol{S}_{\iota}=3aS\mathbf{R}\vect{m}$, we find to second order in $\boldsymbol{\mu}_s$ and the out-of-equilibrium quantity $\vect{m}$
\begin{equation}
	H_{s,u}= 3 aS \lambda_{r}\boldsymbol{\mu}_{s}\cdot \vect{m}. 
\end{equation}

Consequently, the action can be written as $\mathcal{S}= \int {\rm dt} \sum_u [\mathcal{T}_u - H_u - H_{s,u}]$.
Taking the continuum limit, we find the expression
\begin{equation}
	\mathcal{S} = \int dt dA[a_1 \vect{m}\cdot\boldsymbol{\mathcal{V}} - \mathcal{F}_R - \mathcal{F}_m - \mathcal{F}_{\mu_s}],\label{Eq:Action1}
\end{equation} 
where we have introduced $a_1=12\hbar S/ a\sqrt{3}$, $\mathcal{F}_{\mu_s}= g_r \boldsymbol{\mu}_{s}\cdot \vect{m}$ with $g_r = 12 S \lambda_{r} /a\sqrt{3} $, and separated the free energy density of the isolated spin system into two energy contributions arising from $\mathbf{R}$ and $\vect{m}$, respectively.
A minimization of Eq.~\eqref{Eq:Action1} with respect to $\vect{m}$ yields
\begin{equation}
	2 a_2 [1 + \delta_{zi}(\eta_0/ a_2 ) ] m_i = a_1 \mathcal{V}_i + h_{d,i} - g_r  \mu_{s,i}.
\end{equation}
Note that the term $\delta_{zi}(\eta_0/ a_2 )$ only leads to a small correction to $\vect{m}$ on the order of $\eta_0/a_2\sim K_z/J$.
Therefore, we disregard this term in proceeding analysis. Substituting  $ m_i = [a_1 \mathcal{V}_i + h_{d,i} - g_r \mu_{s,i}]/2a_2$ back into Eq.~\eqref{Eq:Action1} and using that 
$\mathcal{V}_i = -(1/2)\epsilon_{ijk} [\mathbf{R}^T \dot{ \mathbf{R}}]_{jk}$, produces the nonlinear sigma model in Eq.~\eqref{Eq:L}. 

\subsection{The dissipative processes}
The following Rayleigh dissipation function models the dissipative processes of the spin system
\begin{equation}\label{General_diss}
	\mathcal{G}=\sum_{i}\int dt\bigg(\frac{\hbar\alpha_{G}}{2} \dot{ \boldsymbol{S}}_{i}^2+ \lambda_{d}\dot{ \boldsymbol{S}}_{i}\cdot(\boldsymbol{\mu}_{s}\times\boldsymbol{S}_{i})\bigg) .
\end{equation}
Here, $\lambda_{d}$ parametrizes the dissipative STT, whereas $\alpha_{G}$ is the Gilbert damping parameter. 

To find an expression for the dissipation in the continuum limit, we consider the contribution $\mathcal{G}_u$ from unit cell $u$ (thus, $\mathcal{G}= \int dt \sum_u \mathcal{G}_u$).
Further, $\mathcal{G}_u= \mathcal{G}_u^{(\alpha_G)} + \mathcal{G}_u^{(\mu_s)}$ is grouped into terms originating from the damping ($\mathcal{G}_u^{(\alpha_G)}$) and the dissipative STT ($\mathcal{G}_u^{(\mu_s)}$). 
Using Eq.~\eqref{S_approx}, we find that $\dot{ \boldsymbol{S}}_{i}^{2}=\dot{ R}_{\alpha\alpha'}\dot{ R}_{\alpha\beta'}n_{i\alpha'}n_{i\beta'}$  to second order in the out-of-equilibrium quantities. 
A summation over the three spins in the unit cell, then leads to 
\begin{equation}
	\mathcal{G}_u^{(\alpha_G)} = \frac{3\hbar\alpha_{G}S^2}{4}\text{Tr}[\dot{ \boldsymbol{R}}^{T}\dot{ \boldsymbol{R}}].  \label{Eq:Gd}
\end{equation}
To second order in $\boldsymbol{\mu}_{s}$ and $\dot{ \mathbf{R}}$, 
the dissipative STT of one unit cell can be written as 
\begin{equation}
	\mathcal{G}^{\mu_s}_{u}= \frac{3}{2} S^2\lambda_{d} \epsilon_{ijk}\mu_{s,\,i}[\boldsymbol{R}\dot{ \boldsymbol{R}}^{T}]_{jk}, \label{Eq:Gs}
\end{equation}
where we have applied $\boldsymbol{S}_{i}\times\dot{ \boldsymbol{S}}_{i}\approx S^2 (\mathbf{R}\boldsymbol{n}_{i})\times(\dot{ \mathbf{R}} \boldsymbol{n}_{i})$ and summed 
over the three sublattice spins. 
In Eqs.~\eqref{Eq:Gd}-\eqref{Eq:Gs}, we have used that $\sum_{k}n_{k\alpha'}n_{k\beta'}=(3/2)\delta_{\alpha'\beta'}$
for the case that the $120^{o}$ ordering is not restricted to lie in the $xy$-plane.
Summing up the contributions from the unit cells and taking the continuum limit produces the following dissipation functional in Eq.~\eqref{Eq:G}. We have defined $\alpha_d=24\hbar\alpha_{G}S^{2}/a^2\sqrt{3}$ and $g_d=12\lambda_{d}S^{2}/a^{2}\sqrt{3}$.

\end{document}